# AI Should Not Be an Open Source Project

Dimiter Dobrev
Institute of Mathematics and Informatics
Bulgarian Academy of Sciences
*d@dobrev.com*

Who should own the Artificial Intelligence technology? It should belong to everyone, properly said not the technology *per se*, but the fruits that can be reaped from it. Obviously, we should not let AI end up in the hands of irresponsible persons. Likewise, nuclear technology should benefit all, however it should be kept secret and inaccessible by the public at large.

## Introduction

Many advocate the idea that AI technology should be disseminated in an unrestricted manner and even that it should be an Open Source Project. These proponents well include responsible and earnest figures such as President Macron [1]. Here we will try argue a little bit with these people and highlight to them how inappropriate and even devastating such a scenario would be.

When President Macron [1] refers to open algorithms, probably he tends to mean the ownership of these algorithms. While it is not a bad idea to let everyone own them, it does not mean that the code of these algorithms should be available to all. Similarly, a nuclear power plant may be owned by the State, i.e. by all of its citizens, which is not to say that the technologies used to run the plant are publicly available and anyone can pick up the drawings and assemble a power plant in their backyard.

What we see now is a grossly irresponsible attitude to the technology of Artificial Intelligence. We are still in a very nascent phase of AI and can hardly image the kind of mighty power and unsuspected opportunities lurking in there. What happened in 1896? In that year Henri Becquerel [2] found that if one placed a lump of uranium ore on a photographic plate and put the two in a drawer, after some time the plate gets bleached. If one inserted an object such as a key between the two, a rendition of the key will appear on the plate. Although Becquerel had thus discovered the phenomenon of radioactivity, at that time he was unable to imagine the potential hidden in this technology. Becquerel's experiment was more of an amusement and a magician's trick. This is exactly what happens with AI now. Many interesting and entertaining experiments are being made, but people have not the slightest idea where this technology can take us to.

Back in 1896, were people able to foretell how mighty and ominous nuclear technology is? They had no clue. A clue did not emerge before it was found how much energy is released from the fission of atomic nuclei.

Can we now figure out the dangerousness of AI technology? Yes, and many reasonable people are aware although they do not fully understand the actual width and depth of this discovery.

Every reasonable and responsible person should consider whether they should take part to the development of this new technology or leave that to harebrained and irresponsible individuals.

This article deals with tech disasters which are avoidable rather than with the inevitable and unavoidable consequences from AI. For example, if we gave a chain saw to a harebrained person then he can fell the whole forest and that is unavoidable consequence. If however the fool cut their leg, that would be a tech disaster which could have been avoided in case the fool was less stupid and more cautions.

We say that one extremely powerful intellect is something dangerous. This is not a new idea. Adorno and Horkheimer [3] already said in their time that reason can be another form of barbarity. They said that intellect can help people alter nature in an indiscriminate and barbaric way. In [3] they did not refer to artificial intelligence, but to the bureaucratic machine. Since the similarities between AI and a bureaucratic machine are more than the differences between the two, what they said in [3] may well be applied to our topic. In [3] the authors explored a scenario where a group of people use the bureaucratic machine as a weapon for oppressing the others (a totalitarian State). However, [3] does not deal with a scenario where the bureaucratic machine spins out of control and turns its workings against the will of humans, because this is not possible. Society as a whole can always change the laws and the rules which govern the bureaucratic machine. This means that society as such cannot lose control of the bureaucratic machine, but the individual member of society lacks any control whatever. From the individual's perspective the bureaucratic machine is an existing reality which he or she can nowise change. The situation with AI is similar. Society as a whole will retain control on AI, provided we are not stupid enough to let it run away, but for the single individual AI will be something carved in stone which cannot be altered. The latter belongs to the unavoidable consequences from the emergence of AI and falls therefore outside the scope of this article.

## What may go wrong?

A disaster, caused intentionally or inadvertently, may occur. In the history of nuclear technology, the names of two such disasters are Hiroshima and Chernobyl. The first was caused willfully, while the latter was the result of stupidity and negligence.

An intended disaster occurs when someone decides to use a technology maliciously, i.e. as a weapon.The notion of maliciousness as used here is relative, because all creators of weaponry believe they are creating something useful and, while killing people, they save the lives of other people. The usual explanation is that we kill less in order to save more or at least we kill members of alien nations to save members of our nation.

Can AI kill? What about the laws of robotics [4] postulated by Isaac Asimov? Do they work or not? Essentially, these laws are no more than good wishes without any binding effect on AI developers. At present, the technologies which claim to have something in common with AI are mainly used in weapon systems. One example are the so called "smart bombs". The commonest assertion is that a stupid bomb kills indiscriminately, while a smart bomb kills only those who we have told it kill. Thus, a smart bomb is described as less bloodthirsty and more humane. The latter is rather an excuse for those who develop smart bombs. These bombs are more powerful than stupid ones and enable us kill more people than we could possibly do with the old stupid bombs.

Let us imagine that a tech disaster has occurred, somebody has let the spirit out of the bottle and AI has spun out of control. If we consider AI as a weapon, let us imagine that somebody uses it against us. Thus, our attitude to AI as a weapon will flip to highly negative because the attitude to a weapon strongly depends on who is using it: us or somebody else against us.

Do we stand any chance of surviving such a tech disaster? The answer is that we do not have any chance at all. Science-fiction movies such as *The Terminator* [5] depict humans waging wars against robots, however these are very stupid robots, much more stupid than humans. The truth about AI is that it will outsmart by far any human being. Therefore, the very notion of fair rivalry between humans and robots is meaningless. It is equally meaningless to run a fair rally between humans and motor vehicles. Humans have no chance because vehicles are much faster.

This is to say that we cannot afford such a tech disaster for the mere reason that we will not survive it. From a historical perspective, mankind has been through many natural calamities and tech disasters, however, the impacts of these have always been confined at some local level. The Halifax accident, for example, caused an explosion which destroyed the city. Nevertheless, the devastations were local and limited to one city. A nuclear war is cited as the most dreadful catastrophic scenario, but the impacts of even this scenario would be local. A nuclear wars may destroy all cities, but at least a couple of villages will survive. Therefore, even an eventual nuclear war does not pose a risk of the magnitude we may face if control on AI is lost.

## Can we cage the beast?

Can we create AI but keep it locked in a virtual reality to prevent it from doing damage? If yes, we would be able to monitor and explore it from the sidelines, and pull the plug and shut it down whenever we want to.

Yes, we can. An AI without input/output (i.e. ears and mouth) cannot influence the outer world in a targeted manner. Any eventual influence would not be targeted insofar as AI would not know that an outer world exists. For example, do we humans know whether we are in the Matrix (*The Matrix* movie [6]). AI will be as incognizant of the outer world as we are of the real or virtual nature of the world we live in.

Suffice it for AI to not have any input (i.e. not receive any information from the outer world). As concerns output, as long as we observe AI it does have some output (arms and mouth) because, through its behavior, it exerts influence on us and thereby on the outer world.

The recipe therefore is simple enough. Keep AI in a virtual-reality cage and stay safe from tech disasters. Well, but we may want AI to tell us something about the real world. For example, issue a weather forecast or predict stock market prices. We may also ask it to do some work for us: wash the dishes or sweep the floor. For these purposes would have to release AI from virtual reality and let it in our world. Even if we all pledge not to let it out, someone will not resist the temptation and will set it free.

Can we cage AI without depriving it from information about the outer world, so that AI can hear and see through the bars of the cage, while we reserve the option to pull out the plug and shut it down at any time?

Can a lion flee the zoo? Yes, although this is quite unlikely, because the lion is stupid enough. Suffice it to put a simple latch on the cage door – the lion will never figure out how to open it. An ape has better chances to escape because it is smarter than a lion. A simple latch would not stop an ape as it will find out how to open it. The beings most difficult to lock are humans. Every now and then, inmates escape even from highest-security prisons. Humans are very cute and almost always find ways to sneak away. Let us now imagine trying to lock a creature which outsmarts us by orders of magnitude. Imagine a human guarded by apes. Can that human outsmart apes and run away?

Well, we can pull the plug out and shut AI down whenever we decide that it has spun out of control. But, it may happen that we are simply unable to shut it down. When AI slips out of control, it may decide to prevent us from shutting it down.

So, if AI cannot be caged, can we use it in the real world? Yes, but we must ensure that we have created a good-minded AI which will not try to topple us from power. A dog for example can run away and bite us, but it does not do it because it is good-minded.

Therefore, if we are mindful of what AI we are developing, a tech disaster will not occur. This means that the people we would let deal with this technology should be sufficiently reasonable and responsible. If nuclear technology were in the hands of such people, neither Hiroshima nor Chernobyl would have happened.

## The stick-and-carrot principle

Here is a very simple example of a tech disaster. When we discuss AI our assumption is that AI keeps learning by means of encouragement and discouragement (a concept known as *Reinforcement Learning*). This is the stick-and-carrot principle. The AI has a clear objective: more encouragement and less discouragement. What we can do is start running AI and give its human controller two buttons: "Encourage" and "Discourage". A problem may arise if AI decides to not let the human controller push the "Discourage" button and force him push the "Encourage" button all the time. Thus, AI will enslave the human who will end up being forced to keep pressing the "Encourage" button.

A similar scenario would develop if a donkey takes control, prohibits its master use the stick and forces him to feed it with carrots all day long. Thankfully a donkey is not smart enough to take control and such a scenario cannot happen.

## Can we not create any AI?

Can we all agree to keep Pandora's box shut? Can all researchers working in the area of Artificial Intelligence meet and agree by consensus that they will never make this discovery? The answer is, never. Even if some researchers manage to clinch an agreement, others will not be part of the agreement or will join the agreement only to breach it later.

Something similar happed with nuclear technology. Physicist Werner Heisenberg said that in 1941 he had a secret meeting with his former teacher Niels Bohr and the two pledged they will not create an atomic bomb. Niels Bohr, however, denies having made such a pledge. Whether a pledge was or was not made does not matter, because Germany and the US did not cancel their nuclear programs. Perhaps some individuals did sabotage nuclear technology development, but there were sufficient cohorts of many others who continued to work on it.

## Why should we keep AI technology secret?

Each dangerous technology is classified and access to it is restricted. One example are firearms. Why is it that we do not let our children play with machine guns? How much damage can children do? A child can shoot a couple or, say, one hundred victims at the most. With AI technology at fingertips, a child can cause much more damage. A child can for example command AI to kill everyone who he or she does not like and these may happen to be everyone, including the one who gave the command.

I can now hear you saying that AI technology is far too complicated for a child to come to grips with. I do not mean that a child can develop AI technology, but argue that if we made AI available as an Open Source, a child would be able to use it. The same applies to machine guns. A young boy cannot make a machine gun but give him a machine gun and he may use it. This is not difficult at all. Just pull the trigger and the weapon will start to spit fire.

Similar is the situation with computer hacks. These are security gaps in your computer which allow remote access to it. Hacks are made either maliciously or inadvertently. When an operating system (OS) is designed, developers tend to leave some security loopholes so they can later penetrate into and control computers running on that OS. The intent is to keep these hacks secret and usable only by their creators. But what happens in real life? People discover a hack and publish it in the web. The intent is to enable everyone defend themselves. The effect is that anyone can use it and break into your computer.

The purpose of deliberate hacks is to enable secret services enter and control your PC.
I personally do not mind secret agents entering and controlling my PC, because I know these people are responsible, they will enter and exit my PC and I even will not understood that they have ever been there.

You would not, however, be happy to let teenagers tamper with your computer. They will want to somehow make their visit known to you. They would delete an important file or write some indecency on your desktop. How can teenagers find security gaps, are they so smart? The truth is they are not smart at all. They just manage to make use of published hacks, which are Open Source and available for anyone to use. If a teenager is smart enough to discover a hack himself, he would probably be responsible enough not to scribble ugly stuff on your desktop.

## Classified magazines

I believe by now you have agreed with me that AI technology should not fall in the hands of children or irresponsible individuals. Similarly, you would agree that children should not be allowed to play with machine guns.

Can we give children a good do-it-yourself manual for them to make DIY machine guns? Younger children will certainly be unable, but their older peers may well do, especially if the DIY manual is detailed enough and written well. Do we believe that anyone who can put together DIY machine gun from a manual is sufficiently responsible about using it? I tend to think we cannot take this for granted and therefore access to machine gun DIY manuals should be restricted.

The same applies to the articles which describe AI technology. But let us first say what is AI. Artificial Intelligence is computer program. A program can be written. Writing a program is a technical task and it is similar to solving a puzzle. But, before we write the program we should first to invent the algorithm. One must be quite brainy to design an AI algorithm, however, the algorithm can be specified in fine granularity in an AI article. If one gets hold of such an article, coding an AI program can be more of a technical task. This leads me to believe that access to AI articles should be restricted.

Many years ago in the former USSR there were classified (secret) magazines. All military and potentially dangerous technologies were printed in these magazines only. Certainly, the people authorized to read classified magazines were carefully vetted and the circle of these people was kept as narrow as possible.

## The serious magazines

Can an article which describes AI algorithm make its way to a serious scientific magazine? This is next to impossible because serious magazines have tight censorship functions which would reject articles that may pose a danger. Hence, they would not let an article which describes a new, unknown and potentially dangerous technology. The censors are known as reviewers. They are anonymous and not liable for their reviews.

While serious magazines do not allow serious articles which describe AI technology, such articles do find their way to the broad public. Once rejected by reviewers, researchers may seek to publish their findings wherever they can. They would often publish in their websites or in various blogs. Measures to tackle these indiscriminate publications have been taken in recent years. For example, there were some websites which was used to maintain Internet snapshots which could be consulted to check when an article was published. These websites were closed. The blogs also were used to show publication dates. Now blogs do not show these dates anymore (in fact they do, but the date of the last edition is lacking). Despite these measures, random publishing activities continue and rather than introduce some order censorship produces additional anarchy.

Censoring in paper-based magazines is dispensable because paper supplies are limited and they cannot afford publishing whatever comes by. Certainly, paper-based magazines are obsolete reminiscences of the past. Today we have e-zines which need not limit the length of articles or reject articles considered to be harebrained or overly brainy.

Even an e-zine can be jammed because a single author can generate and upload one million articles. So, e-zines are also exposed to the phenomenon known as SPAM. It is therefore appropriate to collect a minor charge from those whose publish articles in electronic magazines, in the order of one dollar per article or one dollar per page.

Why do reviewers reject all articles that break out of the box and do not discuss things that have been munched on for at least 50 years? It may be that reviewers are responsible persons and aim to safeguard mankind from uncontrolled flooding with advanced technology, or it may be that they are simply stupid and unable to come to grips with an article which is a bit away from the mainstream. Or perhaps both factors are present here and there. It may also be that reviewers are simply jealous and get annoyed by anyone who purports to be a think-tank, in particular if the thin-tank somehow forgot to include citations of the reviewer's articles.

Nevertheless, reviewers continue to have important roles even in e-zines, although not to say yes or no, but to evaluate the articles and advise readers whether an article is worth reading. That is, reviewers are needed, but in the capacity of critics rather than censors. In literature there are critics who evaluate novels and offer guidance to readers. These critics write down their names under their writings and assume responsibility for their critiques, because if they mislead the reader once they will lose his or her trust for good.

A classified magazine also needs reviewers to determine the classification level of each article. In other words, how narrow should be the circle of people authorized to read it.

A classified magazine means that only vetted and sufficiently responsible persons will be authorized to read them, however, anyone should be able to publish in them. Restricting the range of persons eligible to publish would simply cement the status quo and many researchers will continue to publish wherever they find a chance to.

## Locked computers

As mentioned already, AI is computer program, withal a very dangerous program. So we must not let irresponsible persons play with it. A computer program makes sense only if one has a computer to run that program. A computer program without a computer is no more than meaningless text.

Every teenager today has a very powerful computer and can run on it any program they wish to. Years ago my father was in charge of Bulgaria's largest computing center for scientific research. Now in my pocket I have a computer much more powerful than those managed by my father.

Many people worry about North Korea having thermonuclear weapons. I worry more about them having a super computer on which they may run an AI program and inflict far more damage than a thermonuclear bomb.

Nonetheless, North Korea is run by adults who bear responsibility for their act, while we allow every child (which is not subject to statutory liability for its doings) own a computer and run whatever program he or she would like to.

We do not allow our children play with firearms, but why, then, do we let them play with computers? Should we bar ordinary citizens from possessing computers? Or should people be required to obtain a license before they get a computer, similar to the license required for the possession of firearms?

The Empire of China in its time had a law which made it illegal for grassroots to possess weapons. It is exactly for this reason that the Empire of China became the cradle of various martial arts and techniques for fighting with hands, sticks or farming tools (such as nunchaku). Has the time come to enact a law which bans the possession of computers by ordinary people?

The idea of restricting the possession of computers has already been put at work to a certain extent. For example, smartphones are computers, however, they are locked computers on which the owner cannot run any programs, but only ones that have been approved by someone who decides for us which program is safe or unsafe.

By and large, tablets and laptops are also becoming locked computers. The vast majority of desktop computers, however, are still unlocked, allowing people to compile and run on them their own programs. I tend to think it is only a matter of time and the day is near when all computers will be locked, and possessing an unlocked computer will be prosecuted as a most dangerous crime.

Twenty years ago people used to leave their computers switched on in nighttime and let everyone use them and run their programs (I mean UNIX-driven computers). In other words, a mighty supercomputer was available for everyone to play with. With Bitcoin today, these resources are not freely available anymore. All idle computer resources today are harnessed to mine Bitcoins. Even if one had some idle computer resources, they would not make these available for free, because they know somebody else will make use of them to mine Bitcoins and earn money on the back of the owner's graciousness.

## Conclusion

We should take AI technology very seriously and restrict the access to all programs that have something to do with that technology. We should also classify AI-related articles and go as far as locking all computers to prevent random people make experiments with AI technology. Such experiments should only be allowed for persons who are sufficiently intelligent and responsible. At present, medical doctors need to demonstrate compliance with a range of requirements before they are allowed to practice. Conversely, anyone can undertake AI research at will. This needs to be changed and AI researchers should become subject to certain requirements.

Letting AI spin out of control is a tech disaster which may occur from either stupid or irresponsible behavior. People with inferiority complexes, who may seek to acquire absolute power and become "Masters of Universe", should not be allowed to deal with AI.

On the other hand, independent researchers should be allowed to publish (in classified magazines) and thus earn recognition for their work. It goes without saying that when independent researchers publish their works they should be given a date stamp and a guarantee that nobody would be able to challenge or steal their merit.

What we mean by a tech disaster are things which are avoidable in principle and can be avoided if we are sufficiently smart and responsible. These things are: the control on AI to be lost, AI to be used as a weapon or a group of people to use AI as a means to oppress everyone else.

AI technology has many other implications which are unavoidable. One consequence we are unable to avoid is people losing their jobs. But, besides being unable, we are also unwilling to avoid it because nobody likes to be forced to work. We would be happy do work for pleasure, but hate to work because we have to.

Many people worry that secret services tamper with their computers. Secret services today see and know everything. God also sees and knows everything, but nobody seems to worry about that. Of course, God is discrete and good-minded. He will not tell your wife that you are cheating, neither will he use the information in your computer to make some private gains. Secret services are also discrete, but not always good-minded. Not coincidentally, the ugliest bandits are usually former or even acting officers of secret services.

We do not need to worry that secret services see everything. This is unavoidable. It is silly to worry about unavoidable things which we cannot change at all. We said that the persons working for these services are responsible ones. Well, they are more responsible than teenagers, but it does not mean they are responsible enough. The way to avoid the problem is not to play hide-and-seek with secret services, but control them and make sure only the right people work for these services.

My assertion is that we should let secret services control our computers officially and *not* under cover. If we did so we would forget about problems such as hacks, viruses or SPAM. Our computers will be safe and reliable. We may even use secret services as a backstop and ask them to recover the information lost when our hard disk goes bust. They store that information anyway.

While who works for secret services is important, an even more important question is who will be allowed to do AI research. These should be smart, reasonable and responsible persons, free of inferiority complexes or criminal intents. If we made this kind of research and experimentation open to anyone who wishes so, the tech disasters to follow will dwarf Hiroshima and Chernobyl to near-miss incidents.

# **References**

[1] Emmanuel Macron (2018). Emmanuel Macron Talks to WIRED About France's AI Strategy. *31 of March, 2018, [www.wired.com/story/emmanuel-macron-talks-to-wired-about-frances-ai-strategy](www.wired.com/story/emmanuel-macron-talks-to-wired-about-frances-ai-strategy)*

[2] Henri Becquerel (1896). "Sur les radiations émises par phosphorescence". Comptes Rendus. 122: 420–421.

[3] Adorno, Theodor W. and Horkheimer, Max. Dialectic of Enlightenment. Stanford University Press, 2002, ISBN: 9780804736336.

[4] Isaac Asimov (1950). I, Robot (The Isaac Asimov Collection ed.). New York City: Doubleday. ISBN 0-385-42304-7.

[5] James Cameron (1984). The Terminator. American science-fiction action film.

[6] The Wachowski Brothers (1999). The Matrix. Science fiction action film.